\documentclass[graybox]{svmult}

\usepackage{type1cm}        
\usepackage{makeidx}         
\usepackage{graphicx}        
\usepackage{multicol}        
\usepackage[bottom]{footmisc}
\usepackage{epstopdf}

\usepackage{newtxtext}       %
\usepackage[varvw]{newtxmath}       

\usepackage{comment}

\makeindex             

\begin{document}

\title*{Non-Markovian and Collective Search Strategies}
\author{Hugues Meyer and Heiko Rieger}
\institute{Hugues Meyer \at Saarland University, Theoretical Physics and Center for Biophysics, Campus E2 6, 66123 Saarbr\"ucken, Germany, \email{Hugues.Meyer@uni-saarland.de}
\and Heiko Rieger \at Saarland University, Theoretical Physics and Center for Biophysics, Campus E2 6, 66123 Saarbr\"ucken, Germany; Leibniz Institute for New Materials INM, Campus D1 1, 66123 Saarbr\"ucken, Germany; 
\email{Heiko.Rieger@uni-saarland.de}}
%
\maketitle

\abstract{Agents searching for a target can improve their efficiency 
by memorizing where they have already been searching or by 
cooperating with other searchers and using strategies that
benefit from collective effects. This chapter reviews such 
concepts: non-Markovian and collective search strategies. We start 
with the first passage properties of continuous non-Markovian 
processes and then proceed to the discrete random walker 
with 1-step and n-step memory. Next we discuss
the auto-chemotactic walker, a random walker that produces a 
diffusive chemotactic cue from which the walker tries to avoid.
Then ensembles of agents searching 
for a single target are discussed, whence the search efficiency
may comprise in addition to the first passage time also metabolic costs.
We consider the first passage properties of ensembles of chemotactic 
random walkers and then the pursuit problem, in which searchers 
(or hunters / predators)
see the mobile target over a certain distance. Evasion strategies of 
single or many targets are also elucidated. Finally we review 
collective foraging strategies comprising many searchers and many 
immobile targets. We finish with an outlook on future research directions
comprising yet unexplored search strategies of immune cells and in swarm robotics.}

\section{Introduction}

Understanding search processes, either solitary or collectively, is important in many disciplines like biology, physics computer science, search and rescue and robotics. A review about animal foraging strategies can be found in the books by Sumpter \cite{sumpter} or Viswanathan et al. \cite{viswanathan}. In this chapter we will focus on the question: How can a single searcher use his memory to optimize the search process and how can multiple searchers act collectively, by communication or interaction, to optimize the search process.

\section{Non-Markovian Search Strategies}

Most dynamical processes are not Markovian: their further evolution depends on their past history and not only their instantaneous state. More precisely, the time-evolution of a small set of observables of a system depends {\it a priori} on the values they have previously taken. Of course, any system whose microscopic dynamics is Markovian can be considered as Markovian by simply using the microscopic degrees of freedom as variables of interest. Less extremely, theoretical methods grouped under the name of Markovian embedding procedures have been developed to find sets of observables that are guaranteed to be Markovian. However, the observables derived in this way are not necessarily relevant variables in the context of search processes, where one is mostly interested in the time-evolution of the positions of searchers, which can hence be described as non-Markovian variables.

\subsection{Non-Markovian random walk models}

Many natural systems are equipped with memory and use it to optimize their search efficiency. It can take various forms, either by explicitly remembering the past locations, or by modifying the environment to leave clues of a previous passage. Self-interacting random walks e.g. on a lattice can be defined by nearest neighbor jump rates that 
depend on a weight function $w_i(n)$ that depends on the number
of times $n$ that the walker has visited the site $i$ before.
Introducing a "potential" $V(n)$ by $w(n)=\exp(-V(n))$ one can 
interpret the process as representing a walker depositing a signal 
that, in turn, modifies the local energy landscape $V$ experienced by the walker. In contrast to the auto-chemotactic walker - involving a diffusing chemical clue - discussed below the deposited signal is here assumed to be static and permanent. $V(n)$ could be positive (attraction) or negative (repulsion), and $V(n)=-\beta\theta(n)$ with $\beta\to\infty$ 
corresponds to the self-avoiding random walk [refs].
Such self-interacting or reinforced random walks have been considered in 
\cite{amit,peliti,ottinger,toth,sapozhnikov,pemantle,foster,ordemann,davis,dussutour,boyer,kranz2016,kranz2019}.
Another prominent example for non-Markovian random walks is the elephant walk \cite{schuetz}.

First passage time properties of reinforced random walks have been considered in \cite{barbier} and for non-Markovian polymer reaction kinetics in \cite{guerin}. A general formalism to analyze 
first-passage time properties of non-Markovian processes based only on the properties
of the time dependent mean square displacement has been presented in 
\cite{guerin-nature}. In this chapter we are mainly interested on search strategies, which are specific choices of the parameters of the random walk, and their implications on the mean first passage times (MFPT) - with the ultimate goal to optimize them. This has explicitly been done for random walks with n-step memory and the auto-chemotactic walk.

\subsection{Lattice random walk with 1-step memory}

Many discrete-time models of random walkers are defined such that particles account for their last previous step to determine their next move. Before even mentioning first-passage properties, we should first emphasize that the dynamics of such systems is already significantly impacted by the 1-step memory. In ref. \cite{sadjadi2015persistent}, Sadjadi et al. studied a model where a particle chooses at each time step either to continue in the same direction with a certain probability, or to take a turn with an angle drawn from a probability distribution $R(\theta)$. The diffusive properties of such a particle were analytically calculated and shown to depend in a non-trivial way on $R(\theta)$. Most importantly, the sign of the quantity $\mathcal{R} = \int_{-\pi}^{\pi} d\theta \cos \theta R(\theta)$ was shown to determine whether the dynamics is super-diffusive ($\mathcal{R}>0$) or sub-diffusive ($\mathcal{R}<0$). 

As dynamical properties of searchers greatly influence their first-passage time properties, it is reasonable to expect short-term memory to significantly impact search efficiency. 
In ref. \cite{tejedor2012optimizing}, Tejedor et. al have shown that searchers with short-term memory can already minimize the typical time to find a randomly located target by a substantial amount. The model consists of a random walker evolving on a $d$-dimensional cubic lattice of length $L$ whose probability to continue in the same direction as the one of the previous step is favoured to all other directions. Noting $\epsilon$ the difference between the favoured direction and all other directions, it is shown that the mean first-passage time to reach a randomly located target, in a system with periodic boundary conditions can be calculated analytically as a function of $\epsilon$. Most interestingly, this function is shown to admit a minimum that depends on the system size. In particular, in the limit of large systems, it is shown that the optimal MFPT scales proportionally to $V=L^d$ while it scales as $V\ln V$ for a blind random walk.

Remarkably, optimal search strategies of agents that are trained
by machine learning turn out to resemble closely a persistent 
random walk: In \cite{munozgil} the searcher was a lattice 
random walker equipped with a counter that memorized the number of consecutive straight steps that the searcher performed
before the current time. Reinforced learning was applied to learn
the optimal probability to turn left or right conditioned on the 
number of consecutive straight steps $n$ before, with the result being 
a sigmoidal function of $n$ jumping from zero to non-zero at a specific $n_{\rm opt}$. This search strategy is essentially equivalent to a 
persistent random walk with an optimal persistence length.
Similarly, in \cite{kaur} the searcher was an active Brownian 
particle that can switch between to mobility modes defined by 
zero or non-zero activity, corresponding to zero or non-zero P\'eclet 
number and hence between zero and non-zero persistence.
Reinforced learning techniques were applied to learn the optimal
time intervals to switch between the two mobility modes. It turned
out the the MFPT to find a randomly located target is minimized by
a specific switch rate giving rise to a specific effective (large scale)
persistence length.

\subsection{Lattice random walk with n-step memory}

As memory becomes longer, one can expect a searcher to be more and more efficient to reach a target as it can in principle avoid more easily to return to previously visited locations. In any system there is however a strict lower bound on the mean first-passage time to hit a randomly located target. In a lattice setting for example, the most efficient search strategy for a single walker consists in scanning each lattice site one by one without returning twice on any site. In such a case, the mean first-passage time is given by $T_\infty = V/2$ where $V$ is the total number of sites.  One can therefore wonder about the gain in efficiency using {\it long-term} memory and ask about the number steps in memory required to reach a MFPT as close as possible to the strict lower bound $T_\infty$. 

This question was raised and answered in ref. \cite{meyer2021optimal} by Meyer and Rieger, who formalized in the following way. Consider a walker on a $d$-dimensional lattice who decides where to jump at the next step based on the knowledge of its $n$ previous steps noted $\mathbf{e}_{i_0}, \cdots, \mathbf{e_{i_{n-1}}}$. Formally, it jumps to position $r+\mathbf{e}_k$ from position $\mathbf{r}$ with probability $p(\mathbf{e}_k | \mathbf{e}_{i_0}, \cdots, \mathbf{e_{i_{n-1}}})$. Using a backward master equation, it is possible to show that the mean first-passage time to reach a randomly located target can be calculated exactly from the conditional probabilities $p(\mathbf{e}_k | \mathbf{e}_{i_0}, \cdots, \mathbf{e_{i_{n-1}}})$, provided periodic boundary conditions. This implies that the mean first-passage time can be minimized by optimizing the values of these probabilities. Such an optimization problem can quickly become computationally expensive as the number of variables scales as $z^n$, where $z$ is the coordination number of the lattice (e.g. $z=4$ for a square 2-dimensional lattice) , and the critical step in computing the mean-first passage time consists in inverting a matrix of size $z^n\times z^n$. Hence,as $n$ gets larger, not only the number of variables grows exponentially, but also the computing time to evaluate the objective function. Since, the cases $n=1,2,3$ were numerically optimized and show that for $n>1$, the optimal strategies consist in the concatenation of blocks of steps of length larger than $n$ which can be constructed from the knowledge of the last $n$ steps. In particular, these blocks are not mirror-symmetric, indicating that the left-right symmetry should be broken in order to be efficient in the search. Finally, the optimal strategies lead to a mean first-passage time equal to $V$ for $n=1$, $3/4$ for $n=2$ and $2/3$ for $n=3$, as $V\to \infty$, suggesting the conjecture that the optimal MFPT for an arbitrary value of $n$ scales as $\left(1+\frac{1}{n}\right)\frac{V}{2}$.

\subsection{Autochemotactic Walker}

A natural realization of systems with explicit environmental memory is observed in organisms leaving chemical trails along their path. While the deposition of pheromones by ants is often mentioned as the typical example of such phenomenon, trail-leaving systems are observed in many other contexts and at various scales. At the micron scale, some immune cells as well as a few amoeba species are known to perform auto-chemotaxis, where the cue is produced and emitted by the particles themselves. While the theoretical description of auto-chemotaxis in the continuum for a large number of particles has been introduced in the 1970's via the Keller-Segels equations \cite{keller1971model}, more recent articles have investigated particle-based models in order to study the dynamics and collective behaviours of trail-leaving systems on the level of individual particles. 

The overall dynamics of a trail-leaving particle highly depends on the behaviour of the trail itself. In ref. \cite{kranz2016}, Kranz et al. considered a model of an active brownian particle leaving a finite-width trail which neither diffuses nor decays, which can be considered as an infinite memory timescale. In addition, the trail exerts a repelling force on the particle itself by favoring its director vector to be perpendicular to the trail. The effective dynamics of the particle, namely its effective rotational and translational diffusion constants, is then mainly determined by the quantity $\Omega \tau = k\chi /\pi R v_0^2$, where $k$ is the deposition rate of the trail, $\chi$ is the sensitivity of the particle to the trail and $v_0$ is the active speed of the particle. In particular, a phase transition is observed at $\Omega\tau > 2$, where the rotational diffusion constant diverges and the system transitions from a diffusive behaviour to a state where the particle performs very confined loops and is practically trapped by its own trail in a small region of space. This regime if of course unfavorable for a search process as space exploration is considerably hindered.

\begin{figure}
    \centering
    \includegraphics[width=.42\linewidth]{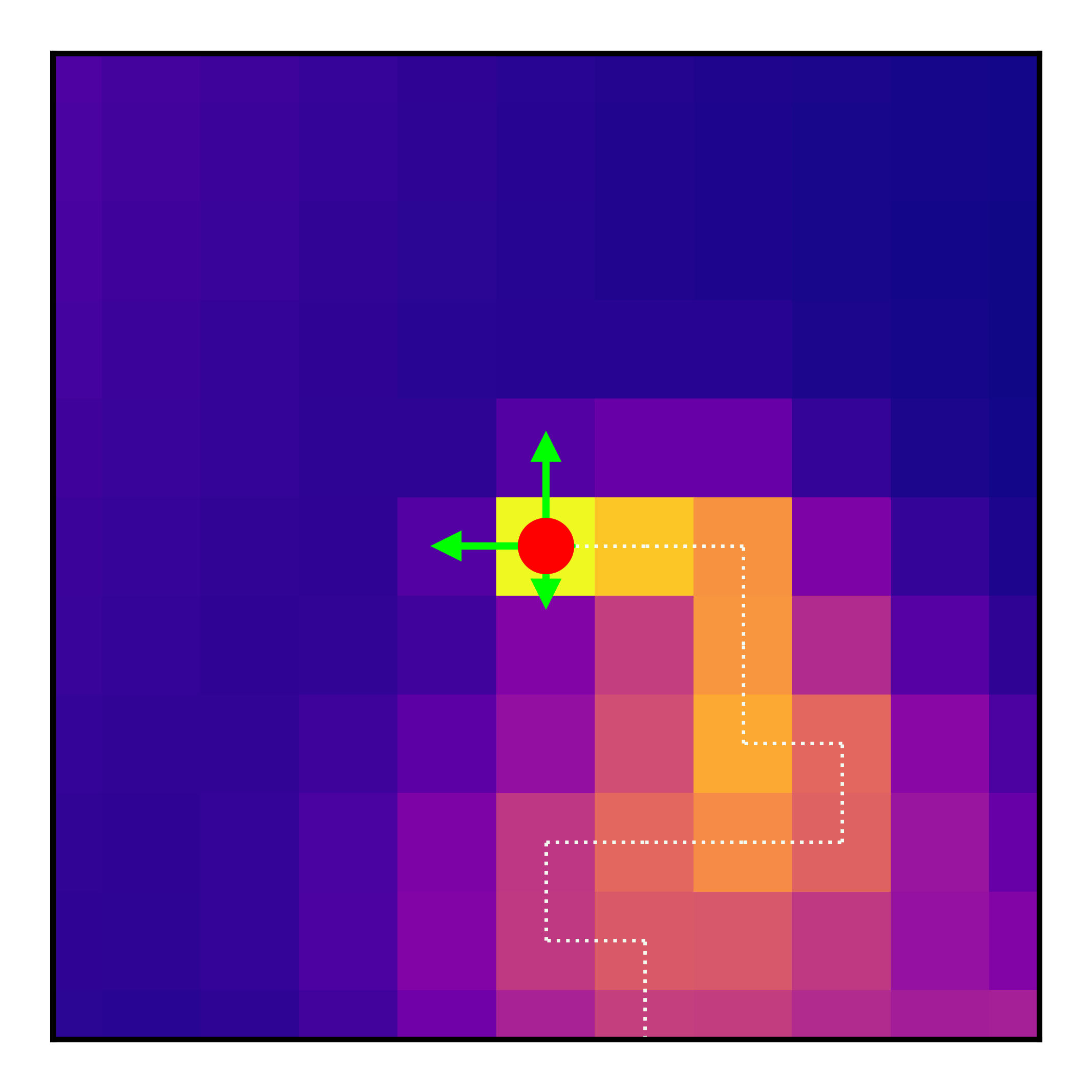}
    \includegraphics[width=.56\linewidth]{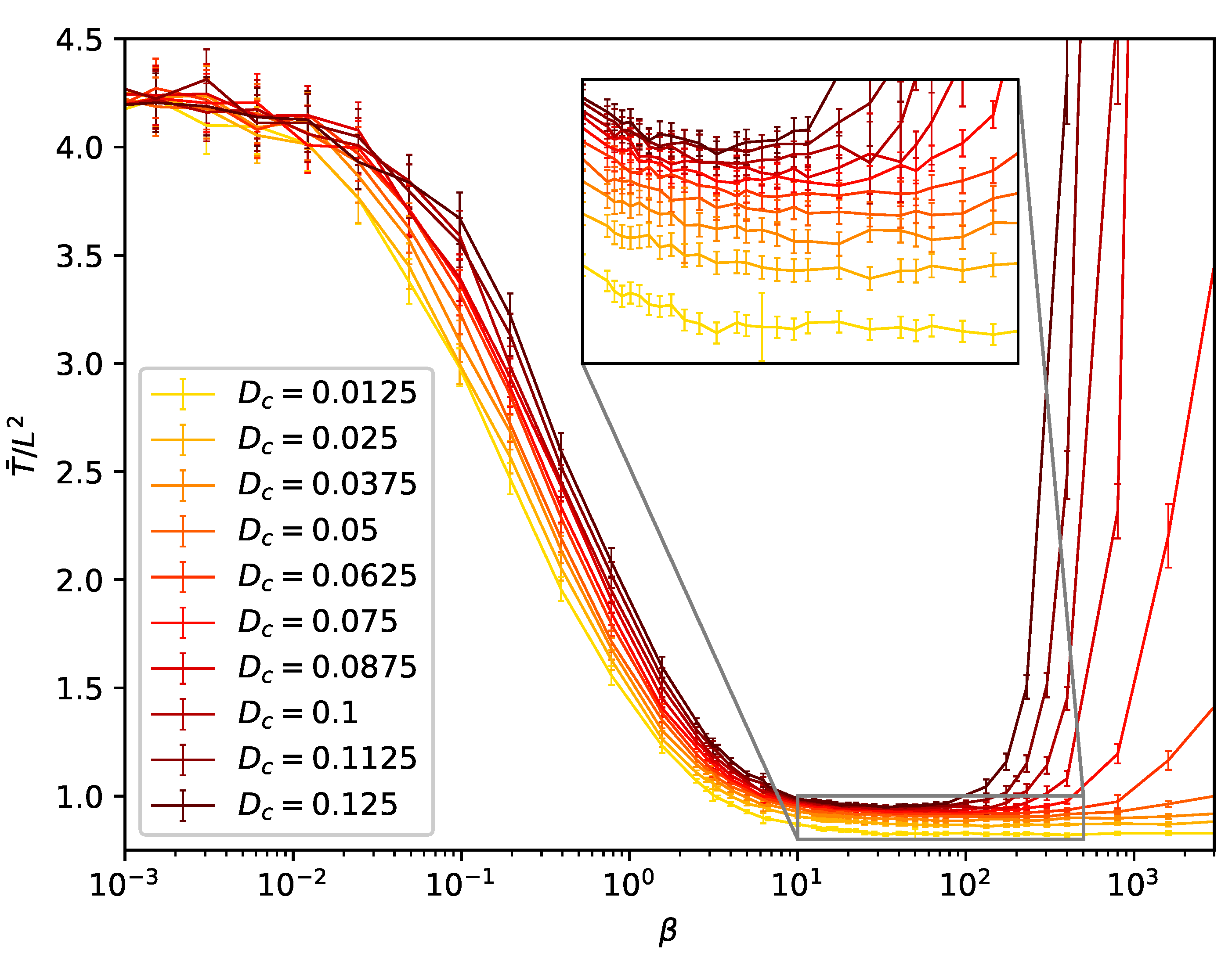}
    \caption{Snapshot of an auto-chemotactic walk as defined in \cite{meyer2021optimal}. Left panel: The red circle indicates the position of the walker while the dotted line tracks its path and the color gets darker for sites visited further in the past. The background color codes for the concentration field, which determines the probability to jump to one of the neighbouring sites at the next step as indicated by the size of the green arrows. Right panel: 
    Mean first passage time $\overline{T}$ of a single auto-chemotactic walker to reach a randomly located target, normalized by the system size, as a function of the coupling constant $\beta$ between the 
    walker and the chemotactic field for various values of the diffusion constant $D_c$ of the chemotactic field. From \cite{meyer2021optimal}.}
    \label{fig:autochemotactic}
\end{figure}

If the chemical cue starts diffusing and/or decaying, one enters the realm of auto-chemotactic particles, a time- and space-continuous model of which was introduced in \cite{taktikos2011modeling}. There, the particle is again modeled as a point-like active Brownian particle whose velocity unit vector experiences a torque generated by the chemotactic field, i.e. the gradient of the concentration field. In the weak chemotactic coupling limit, the effective dynamics of the walker is simply diffusive and evolves according to a Langevin equation parameterized by an effective diffusion constant. For a strong attractive coupling (i.e. for a particle attracted by the trail), circular trajectories are also observed, but not in the case of chemo-repulsion. A similar dependence of the dynamics of an auto-chemo-repulsive particle was reported in ref. \cite{meyer2021optimal}, where a simpler lattice model was introduced, c.f. Fig. 1a. There, a walker jumps from a lattice site to a neighbouring with a probability favoring sites with low values of concentration field defined on the lattice. Then it then adds a fixed amount to this field at the location it has jumped to. The walker is showed to have a persistence length highly dependent on the diffusion constant of the concentration field, which effectively acts as a pushing and aligning force on the walker (see Fig. 1a for a snapshot of the emerging concentration). It turns out that that the MFPT has a pronounced 
minimum for a specific coupling constant between the walker and the chemotactic field as shown in Fig. 1b. In \cite{meyer2021optimal,} it
was also shown that the main effect of increasing $\beta$ was to 
increase the persistence length of the auto-chemotactic walker, 
implying that the optimal coupling constant corresponds to an
optimal persistence length.

\subsection{Miscellaneous}

Other non-Markovian searches have been studied in recent literature, often motivated by experimental observation of biological systems. In the realm of systems that remember previously visited locations, it is e.g. observed in some animal species, such as bats, monkeys or even humans, that the probability of visiting new sites depends on the total number of sites already visited, and the probability to choose a specific site among all visited sites depends on the number of previous visits on that site \cite{boyer2014random, vilk2022phase}. It is shown that this dependence can be fine-tuned for the search to allow frequent returns to preferred sites while enabling an extensive scan of all sites of the domain, which leads to a good balance between energy consumption and risk management. On the smaller scale, similar conditional behaviours leading to long-term memory effects are also observed and modeled. In \cite{jose}, the hopping of two species of molecular motors along a couple of microtubules is considered. Each species can travel along a single direction, and each motor can either bind or unbind to a filament, or hop along it, and the probability of binding to a filament changes whether the last motors of that have been on that portion of the filament were of the same species or not. Transport properties and in particular the efficiency of bidirectional transport is shown to be greatly influenced by such a conditional binding probability.

More explicit non-Markovian searches have also been investigated, under the form of Generalized Langevin Equations (GLE) where memory is directly encoded in the equation of motion of the searcher via a memory kernel. While it is known that the overall dynamics, in particular first-passage properties, are highly sensitive to the form of the memory kernel of the GLE \cite{hanggi1985first}, it was shown numerically that exponentially decaying kernels are optimally efficient in the context of composite searches, i.e. searches which alternate between intensive and extensive searches, for spatially correlated targets \cite{klimek2022optimal}.

Finally, searches with environmental memory like the auto-chemotactic walk were introduced in recent years. In particular, the Sokoban walk \cite{bonomo2023loss} considers a walker on a lattice where obstacles are placed. The walker can push the obstacle with a certain probability, provided that there is no other obstacle behind it. An experimental realization of a similar system using bristle robots and light movable obstacles was presented in \cite{altshuler2023environmental} and showed that the effective memory created via the trails of the robot through the obstacles allows for a more efficient search for randomly located target.

\section{Collective Search Strategies}

Multiple searchers will generally find a single target faster than a single one, see \cite{redner,nayak} and the process can be collectively optimized by utilizing interactions or communications between them. The standard, against which any optimization by collectivism should be measured, is the FPT (to reach a single target) distribution of a number $N$ of non-interacting random searchers, which is given by the distribution of the minimum of FPTs of N individual searchers. Typically, one can expect that the MFPT of N searchers is proportional to $1/N$ times the MFPT of a single searcher \cite{redner,nayak}, but the 
ensemble-averaged realization-dependent first passage time 
exhibits quite a non-trivial and sometimes counter-intuitive behavior
\cite{meyer2021optimal}.

A simple strategy for $N$ random searchers to improve their search efficiency for a single or many targets would be to avoid clustering and to achieve a homogeneous distribution over the search area. A theoretical model for random walkers interacting via long-range repulsion (e.g. Coulomb potential) has been proposed in \cite{tani} and it has been found that the search efficiency can indeed be substantially increased by this interaction: the MFPT shows a pronounced minimum for a specific interaction strength, analogous to the charge in the Coulomb interactions. The observation that the MFPT depends non-monotonously on the interaction strength is plausible since a too strong repulsion decreases the effective diffusion constant of each searcher.

\subsection{Chemotactic searchers}

A similar strategy underlies random searchers with chemotactically repulsive interactions \cite{meyer2022alignment}: each searcher produces a chemotactic clue that diffuses into the environment and tells other searchers to turn away. This behavior is modelled by incorporating the gradient of the concentration field of the chemotactic clue into the equations of motion of each searcher: on a lattice by a concentration (difference) dependence of the transition probabilities \cite{meyer2022alignment} and in continuous space by a concentration (gradient) torque in the equation of motion for the angle of, e.g., active Brownian particles (Wysocki and Rieger, unpublished). It has been found that in both cases the MFPT can be substantially reduced by a specific choice of the strength of the interaction between the particle and the chemotactic clue. It should noted that the chemotactic clue can already improve the search efficiency of a single searcher since the self-produced concentration field leads to a preference to move in the same direction, away from the clue, and thus increases effectively the persistence length of the searcher – which has already the potential to improve and optimize the search efficiency \cite{tejedor2012optimizing}. For multiple searchers it has been shown that in addition to the single particle effect the MFPT can be decreased further, a manifestation of a collective effect.

In the aforementioned example the search efficiency measured by the MFPT alone is a monotonous function of the number of searchers, N: simply decreasing with N. This would imply that it would always be better to involve as many searchers as possible to have the most efficient search. From a searchers perspective this would be unrealistic since the cost of a search is not given by the search time alone but usually also contains a metabolic contribution: each searcher consumes energy to move. Incorporating this metabolic cost into the measure for search efficiency leads to a well-defined optimal number of searchers, $N_{opt}$ (Meyer and Rieger, to be published). However, its should be noted that for large enough diffusion constants of the concentration field, auto-chemorepulsive particles can self-organize and form macroscopic bands traveling at constant speed as the particle density increases, due to the effective alignment interaction between particles mediated by the concentration field. Such structures are deleterious for search processes as particles are confined in a small region of space and thus do not scan space efficiently. A detailed analysis of this phenomenon is presented in \cite{meyer2022alignment}, where phase diagrams are reported. 

Note that auto-chemoattraction is also relevant for collective search processes. In fact it can be shown to be an optimal collective search strategy when accounting for various contributions to the search efficiency \cite{pezzotta2018chemotaxis}. Consider a set of $N$ particles evolving according to Brownian dynamics driven by the gradient of a potential $Z$ of the form $Z(x_1,\cdots,x_N) = \zeta(x_1)\cdots\zeta(x_N)$ where $x_i$ is the position of particle $i$. A cost function defined as the weighted contributions of the elapsed time, the kinetic energy and the collisions between particles until all particles have found a target is minimized if and only if the function $\zeta$ and the particle density field satisfy the Keller-Segel equations for auto-chemoattraction. This is particularly relevant in the context of scouting, occurring when a first particle has found the target, stops moving and recruits other particles guided by the cue it produces.

\subsection{Chemotactic searcher-target attraction}

We have discussed so far how a repulsive (auto)chemotactic searcher-searcher interaction can improve search efficiency. In many search problems it is the target that emits a signal or chemical clue received by the searcher, mostly attractive like bacteria searched by macrophages or rabbits hunted by dogs, or repulsive as an evasion strategy of the target.
Typically the searcher, or predator, moves along the steepest gradient of the chemical concentration to ultimately find its ejection source. Likewise the prey (or the target, for example another microbe) “smells” a secreted
chemical from the advancing predator and tries to escape by moving along in the opposite direction of its maximal gradient. For such a situation a discrete chemotactic predator-prey model was proposed in \cite{sengupta}, in which the prey secrets a diffusing chemical which is sensed by the predator and vice versa. Two dynamical states corresponding to catching and escaping were identified and it is shown that steady hunting is unstable. 
A subsequent work analyzed the efficiency of such a chemotactic pursuit and comparing blind search with temporal and spatial gradient sensing \cite{metzner}. Although strategies based on the temporal or spatial sensing of chemotactic gradients are significantly more efficient than unguided migration, a ''blind search'' turned out to work surprisingly well, in particular if the searcher is fast and directionally persistent. This result underpins the importance of the persistent random walk with an optimally chosen persistence length (c.f. section 2.2) as an efficient search strategy in comparison with chemotactic gradient search.

More complex searcher-target interactions emerge for instance in the immune systems, where e.g. T cell migration is essential for T cell responses and an efficient immune response to virus infected or tumor cells. It allows for the detection of cognate antigen at the surface of antigen-presenting cells and for interactions with other cells involved in the immune response. Although appearing random, growing evidence suggests that T cell motility patterns are strategic and governed by mechanisms that are optimized for both the activation stage of the cell and for environment-specific cues \cite{krummel}.

Another example is the neutrophil recruitment from blood to extravascular sites of infectious tissue damage, which is a hallmark of early innate immune responses. Once outside the vessel, individual neutrophils often show extremely coordinated chemotaxis and cluster formation reminiscent of the swarming behaviour of insects. Such a coordinated chemotaxis over several millimeters cannot be achieved by diffusion alone: with a
diffusion constant of $D\approx10^{-10}{\rm m}^2/{\rm sec}$ for small signalling molecules diffusive signaling would take hours which is incompatible with the observed time scales in neutrophil swarming. In \cite{laemmermann} it was revealed how local events are propagated over large-range distances, and how auto-signalling produces coordinated, self-organized neutrophil-swarming behaviour that isolates the wound or infectious site from surrounding viable tissue. These local events constitute a so-called signaling relay: the first neutrophil searcher reaching the target emits a signal that diffuses into the environment. When other neutrophils close by detect this signal they also start to produce the signal and so forth (c.f. Fig.1 in the Supplementary Information of \cite{laemmermann}). Similar swarming behavior has been observed in 
T cells \cite{nino}. The dynamics of such diffusive signalling relays, which are also relevant in developmental biology and microbial consortia,
has been analyzed in \cite{dieterle2020dynamics} and it was shown that it gives rise to signalling waves propagating with a density and diffusion constant dependent velocity $v$. With a grain of salt this propagation velocity can be interpreted as a measure for the efficiency of the observed collective search strategy.

It should be noted that a vast literature exists on chemotaxis among a single group of individuals (i.e. without a searcher-target scenario), leading to agglomeration and pattern formation in the attractive case, mostly based on the Patlak-Keller-Segel model, for which we refer to \cite{painter} for a review.

\subsection{The pursuit problem: Collective chase}

In the pursuit problem the searcher (or hunter / predator) can “see” the target (or prey), which either moves deterministically or stochastically. 
In an analytically solvable model for $N$ non-interacting predators 
hunting a single randomly moving prey it was assumed that the predators
always move in the direction towards the prey, which is assumed to be always visible \cite{bernardi}. 
This is neither the most efficient pursuit strategy for the predator 
nor the most efficient escape strategy for the prey, but this simple model allowed for the analytical computation of the mean capture time. It turns out that depending on the number of chasers, the mean capture time as a function of the prey’s diffusion coefficient can be monotonically increasing, decreasing, or attain a minimum at a finite value \cite{bernardi}. From the hunter’s perspective, possibly more relevant than the mean capture is the energy spent by the hounds to capture the prey (c.f. chemotactic search above). Then an optimal speed and number of chasers exist and depends on each chaser’s baseline power consumption \cite{bernardi}. The extension of the model to the case in which the stochastic prey that sees its hunters was discussed in \cite{su}. The evasion strategy of the target consisted in a deterministic contribution to the stochastic equation of motion, which averages over repulsive forces exerted by the hunter, each weighted by exponential decay with distance from the hunter.

\begin{figure}
    \centering
    \includegraphics[width=\linewidth]{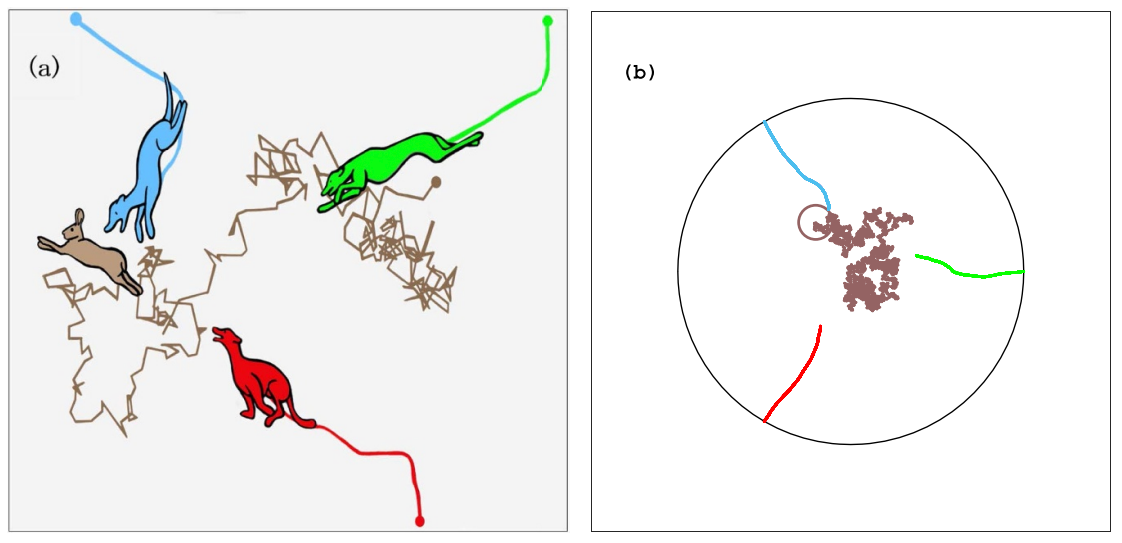}
    \caption{(a) Sketch of the group hunting model studied in \cite{su}.
    (a) The hare/prey diffuses according to simple Brownian dynamics while the hounds/predators run at constant speed and adjust their direction to always aim at the prey. (b) Typical trajectories of the model studied in \cite{su} for three 
    hunters and one target starting from the circular boundary and the 
    circle center, respectively. From \cite{su}.}
    \label{fig:chase}
\end{figure}

To analyze efficient group chase strategies interactions between predators have to be considered (analogously for escape strategies of groups of prey, to be discussed in the next section), which results in a many-body problem that in most cases can only be tackled numerically, see for instance
\cite{vidal,romanczuk,kamimura,angelani,iwama,lin,janosov,surendran}.
An active matter approach \cite{shaebani} was taken in 
\cite{angelani}, where two groups of self-propelled particles were 
considered, each with intra-group alignment as in the Vicsek model for
flocking \cite{vicsek} and repulsion (to avoid collision).
Chase and escape were described by inter-group interactions, attraction (for predators) or repulsion (for preys) from nearest particles of the opposite group. It was found that the capture time depends 
non-monotonously on the effective range of the predator-prey 
interaction, and could be maximized in favor of the prey.

A more biology inspired approach modeling group chase tactics
was taken in \cite{janosov}, considering in particular the challenge
for predator groups to catch a faster prey: 
For the predators or chasers 
the model contained equations of motion describing 
that each predator chases the closest target. The collective chasing strategy included a soft repulsion between the chasers (to avoid 
clustering of the predators emerging frequently when all predators 
move towards the same spot, the location of the prey)
and a prediction of their target’s position. 
The escape strategy of the prey was mainly defined by moving in the free direction that is furthest away from all the chasers within the sensitivity range, but also contained a stochastic component describing 
seemingly erratic escape behavior frequently observed in nature.
A panic parameter was introduced that depends on the distance between the escaper and its nearest chaser and controls this erratic behaviour.
In nature faster prey are frequently captured by predators when 
running into obstacles or confining walls. Consequently the model also contained rules for escape behavior at walls: 
At the wall confining the simulation area, the escaper aligns its velocity to the wall. If the escaper can slip away between the nearest two chasers, then it returns to the field. 

It was found that repulsive interactions between the predators are
important for capture efficiency and can lead to optimal groups 
of chasers that can catch a faster prey. It was also observed that emergent behavior can occur: with certain parameters the chasers have the chance to encircle (encage) their prey. If chasers use the prediction method to forecast their targets' position, their efficiency increases, and this can also overcome great delay times in the chasers reactions to prey movement. 
Finally, erratic escape behavior of prey in the form of stochastic zigzag 
movement can be advantageous for the prey, when there is delay in the system \cite{janosov}, which brings us to the discussion of target evasion strategies - in particular collective evasion strategies.

\subsection{Evasion strategies}

A variant of the pursuit problem in which the predators have only limited sight was studied in \cite{oshanin}: here the hunters perform independent random walks on a square lattice with V sites and start a direct chase whenever the prey appears within their sighting range. The prey is caught when a predator jumps to the site occupied by the prey.  When the prey is blind, i.e. it has no information on predators’ actual positions (its sighting range is zero), the best strategy is to stay still \cite{bray,moreau1,moreau2}. Contrastingly, in \cite{oshanin} the efficiency of a minimal-effort evasion strategy was analyzed according to which the prey tries to avoid encounters with the predators by making a hop only when any of the predators appears within its sighting range; otherwise the prey stays still. It was shown that if the sighting range of such a lazy prey is equal to 1 lattice spacing, at least 3 predators are needed in order to catch the prey on a square lattice. In this situation, a simple asymptotic relation $\ln P_{ev}(t) \propto (N/V)^2 \ln P_{imm}(t)$ between the survival probabilities of an evasive and an immobile prey was established. Hence, when the density $\rho = N/V$ of the predators is low, $\rho\ll 1$, the lazy evasion strategy leads to the spectacular increase of the survival probability \cite{oshanin}.
A single predator hunting a herd of prey, both with limited sight, 
was also considered in \cite{schwarzl}, including explicitly the self volume of the prey restraining their dynamics on the lattice.
It was found that the prey's sighting range dominates their life expectancy and the predator profits more from a bad eye-sight of the prey than from his own good eye sight.

Collective evasion strategies, i.e collective behavioral patterns 
of groups of prey when attacked by a predator,
can be observed in animal swarms, herds and flocks, with the primary function to protect members when attacked by predators. One way to reduce the predator’s chance of making a successful kill is to confuse the predator, which may be accomplished by collective evasion behavior. In \cite{zheng2005behavior} such a behavioral pattern was modeled by self-propelled particles (in discrete time) choosing their direction of motion 
to a) achieve flocking by rules for medium range alignment, short-range collision avoidance and long range attraction and b) escape rules that 
are activated when a (single) predator is in sighting range. These escape rules can either be selfish or cooperative, the latter implying a "patience" zone in which escape is not activated although the predator is in sight. The (single) predator is also modeled as self-propelled particle and chooses at each time step one prey randomly within its attack zone having a fan shape whose center axis is in the direction from the predator to its previous target and whose spread angle is fixed. The predator tries to approach directly the target at each moment, but its ability to to change its movement direction is in general inferior to the ability of prey.
It was shown that the mean first capture time is maximized for a specific value of the width of the "patience" zone, underpinning the efficiency of cooperative evasion strategies. 
\begin{figure}
    \centering
    \includegraphics[width=\linewidth]{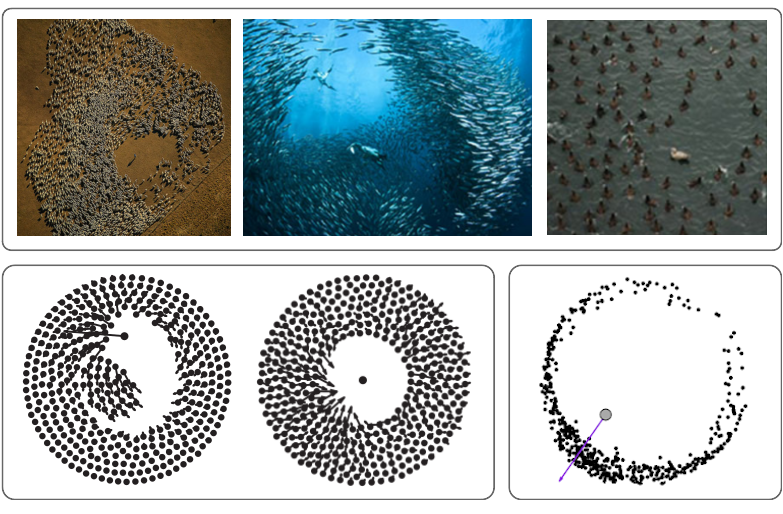}
    \caption{Upper panel: Pictures of the evasion strategy of
    groups of sheep (left), sardines (middle) and ducks (right)
    in the presence of a predator (always in the center of the 
    emerging "donut" as observed in nature. From \cite{chen}.
    Lower panel: Simulation snapshots of a group evasion strategy 
    model studied in \cite{chen}. Left: first-order model (i.e. overdamped equations of motion for the particle velocities) with two different predator-pray attraction strength. Right: second-order model (i.e. equations of motion for the accelerations). From \cite{chen}.}
    \label{fig:evasion}
\end{figure}

A more minimalistic approach, but essentially in the same spirit, was taken in \cite{chen}: The movement of prey particles and a single predator were described by (over-damped) equations of motion in which the velocity changes were determined by repulsive and attractive forces (decaying algebraically with distance). Prey-prey interactions comprised an attractive long-range part (leading to flocking) and a repulsive short-range part (avoiding collisions), prey-predator interaction was repulsive and predator-prey interaction was an average over attractive forces from each prey. The model predicts the formation of a stable donut-like arrangement of the prey positions around the predator in certain parameter ranges, an emergent collective behavior also observed in nature (c.f. Fig. 3). The predator can be ‘trapped’ at the centre of the prey swarm while the prey forms a concentric annulus where the repulsion exerted by the predator cancels out owing to the symmetry. If the predator strength $c$, the coupling parameter between the average predator-prey attractive forces and the velocity change of the predator, is sufficiently small, the swarm will escape the predator. If $c$ is increased past a threshold,
the predator becomes more ‘focused’ and less ‘confused’, resulting in ‘chasing dynamics’ which can lead to very complex periodic or chaotic behaviour \cite{chen}. Thus for a ‘weak’ predator, the swarm is able to escape the predator completely.

\subsection{Collective Foraging}

In contrast to "hunting" discussed in the previous sections, which involved mobile prey, we focus here on collective "foraging", meaning many searchers and many immobile targets, like plant food for herbivores. In contrast to 
the mathematical problem for $N$ searchers and $M$ targets \cite{palyulin2017comparison, khadem2021search}, 
foraging involves the consumption / removal 
of a target once it is found and usually also the renewal of targets,
such that the target density remains approximately constant. 
Thus, the foraging process can reach a stationary state, in 
which the foraging efficiency, measured by targets found per time,
is of interest. In particular when the distribution of targets is
"patchy" (i.e. randomly distributed high density clusters of targets
in an environment with otherwise low target density) it appears 
plausible that communication between the searchers could be useful.

This hypothesis was scrutinized in \cite{martinez2013optimizing}
where a model of a group of interacting foragers was introduced
to study the optimal strategy to search for heterogeneously
distributed targets. In the model, foragers move on a square lattice containing immobile but regenerative targets. At any instant, a forager is able to detect only those targets that happen to be in the same site. However, the foragers are allowed to have information about the state of other foragers. A forager who has not detected any target walks towards the nearest location, where another forager has detected a target, with a probability $\exp(-\alpha d)$, where $d$ is the distance between the foragers and $\alpha$ is a parameter characterizing the propensity of the foragers to aggregate. The model reveals that neither overcrowding ($\alpha\to0$) nor independent searching ($\alpha\to\infty$) is beneficial for the foragers. For a patchy distribution of targets, the efficiency is maximum for intermediate values of $\alpha$ \cite{martinez2013optimizing}.

A continuum model for interacting foragers was proposed in \cite{bhattacharya2014collective}: The motion of the foragers was described 
by Langevin equations for the position, with two drift terms: one derived from a environmental quality function (amount of grass, prey, etc.) and one derived from an interaction potential between the foragers. The latter just spreads with a Gaussian kernel with width $\sigma$ the local information that a forager has about the environmental quality. To illustrate the effect of this inter-forager communication a theoretical landscape quality function consisting of three non-normalized Gaussian functions (centered at different target positions) was considered and the mean first arrival time of foragers at the targets was evaluated. It was shown that 
the so-defined search could be minimized for a specific value of the interaction or communication range $\sigma$: the number of individuals from which a given searcher receives a signal will typically increase with the interaction range. When this scale is too small, individuals receive too little information (no information when $\sigma=0$), and thus exhibit low search efficiency. Similarly, interaction ranges that are too large lead to individuals being overwhelmed with information from all directions, also resulting in inefficient searches \cite{bhattacharya2014collective}.
As an application the model was applied to the acoustic communication among Mongolian gazelles, for which data are available, searching for good habitat areas. The key observation of the model and the application is that, in general, intermediate communication distances optimize search efficiency in terms of time and quality.

\section{Summary and outlook}

As soon as many searchers and / or many targets are involved in a target problem, collective search or evasion strategy are advantageous – either for the searcher or the evader. A few fundamental facts about collective searches are known and have been analyzed quantitatively in the past, as we reviewed here: for many agents searching randomly (i.e. blindly) one or many randomly distributed targets a repulsive interaction between them, for instance via chemotaxis, is advantageous; in case random searchers are supposed to swarm to a specific randomly located spot over long distances (e.g. neutrophils to a wound) a signaling relay is useful to overcome the limits of diffusion; for group hunting (i.e many searchers with limited visibility) repulsion between the hunters and prey trajectory prediction is advantageous; for group evasion (i.e. escape behavior of prey seeing the predator) attractive interaction between prey and a mixture of deterministic and random movement in the presence of a predator is advantageous; for group foraging searchers should communicate the local target density to others nearby.

Nature is of course more sophisticated than this and developed many more collective search strategies which are largely unexplored. For instance, the immune system is one arena where searching (and killing) is of outmost importance. T cells change their migratory behavior in different environments – like in lymph nodes or tissue \cite{krummel}. There are striking similarities between strategies of immune cells to find pathogens and strategies of ant colonies to forage for food and it has been hypothesize that the ability of searchers to sense and adapt to dynamic targets and environmental conditions enhances search effectiveness through adjustments to movement and communication patterns \cite{moses}. Other immune cells, like T helper cells, B cells, dendritic cells, macrophages etc, are involved in an efficient immune response, too, interacting with each other and altering migratory and secretory patterns.

Inspired by such natural examples a plentitude of complex search strategies could be imagined: searchers adapting their migration behavior to a spatially and/or dynamically changing environment or target distribution; different types of searchers communicating in different ways and with different ranges; mobile or immobile bystander agents that influence or relay information about target distribution or type, and much more. Similarly sophisticated escape strategies could be imagined, like switching between flocking, group splitting and dispersion to confuse an approaching predator, long range communication or the involvement of bystander providing information about predator movements, etc. 
Moreover, one could expect significant advantages of search and
escape strategies that involve memory of the searchers or the moving
targets (the escapers) like in the non-Markovian search processes that 
we reviewed in the beginning of this chapter.

More complex search and hunting strategies than those reviewed here have been developed in social animals during evolution. Four levels of increasing complexity characterize group-hunting \cite{vincent}. The lowest level of complexity, similarity, is characterized by individuals directing their behaviors at a single target, as we reviewed in section 3.3.; however, these actions lack temporal and spatial coordination. When the behaviors of the group align temporally, it is known as synchrony. Coordination represents a further increase in complexity as individuals relate their actions temporally and spatially to the actions of the other group members. Finally, the highest level of complexity is collaboration. When collaborating, individuals not only coordinate their behaviors, but also perform a variety of complementary behaviors directed at the same target, which may require understanding the others’ role in completing the goal. Examples for the highest levels of complexity in group hunting among social animals have been reviewed in \cite{vincent}. 

Swarm intelligence observed in nature often stimulated the design of analogous strategies in artificial systems like robots \cite{bonabeau,brambilla}. For instance, sophisticated foraging strategies were implemented in swarms of unmanned aerial vehicles (UAV) or drones, which have become more and more accessible and are now increasingly utilized across a variety of domains, including exploration of unknown territories and monitoring of hazardous environments. In its simplest form, a swarm of UAVs is characterized by a large number of homogeneous individuals (agents) with local communication, sensing and actuation capabilities and its swarm behavior can be designed by employing three main conceptual frameworks \cite{brambilla}: (i) collective decision making, in which UAVs aim at achieving consensus on a collective strategy, in order to maximize the performance of the swarm; (ii) spatially organized behavior, in which UAVs aggregate according to a set of spatial constraints to arrange themselves following specific patterns, chains, or structures; (iii) navigation-based behavior, in which each UAV follows a set of simple behavioral rules to steer the whole swarm and cooperatively explore an environment. The latter is mostly used for search and coverage task and is frequently designed by exploiting biologically inspired approaches \cite{alfeo}.

In \cite{alfeo} swarm behavior of UAVs was modeled considering and comparing two different paradigms, namely biological behavior, which mimics social animal metaheuristics, and computational behavior, which considers enhancements exploiting UAVs information technology. For instance, computational flocking involved different rules for movement in three different zones: “separate” in the near zone (corresponding to short range inter-searcher repulsion), “align” with neighbors in the medium zone (corresponding to the Vicsek model) and “cohere” in the far field zone (corresponding to long-range attraction). Note that similar movement rules
were applied to groups of prey/escapers in some of the models for evasion strategies reviewed in section 3.4.
Computational “stigmergy” (meaning agents leaving trails in the environment to coordinate other’s movement, like chemotaxis) involves “digital pheromones”, which are maintained in a virtual space, called pheromone map, and can have an instant diffusion, to immediately propagate the environmental information to nearby UAV. Pheromones models can include both attractive and repulsive pheromones, to mark detected targets and visited areas, respectively. The various possible strategies are then usually optimized by using evolutionary algorithms or machine learning.

Finally, we would like to mention that chemotaxis and swarming is not only ubiquitous in biological systems but has recently been found in synthetic active matter, like active colloids, too \cite{liebchen2017,liebchen2018, grauer2020}. This offers the possibility to explore collective search strategies experimentally, which would pave the way for novel applications,
as for instance in (nano)medicine.


\end{document}